\documentclass[%
aip,jcp,numerical,
amsmath,amssymb,
reprint,%
]{revtex4-2}

\usepackage{graphicx}
\usepackage{subfigure}
\usepackage{amsmath}
\usepackage{amssymb}
\usepackage{tikz}
\usepackage{dcolumn}
\usepackage{bm}
\usepackage{blindtext}
\usepackage{listings}
\usepackage{color}
\usepackage{physics}
\usetikzlibrary{quantikz}
\usepackage[colorlinks,linkcolor=blue,anchorcolor=blue,citecolor=blue]{hyperref}
\usepackage{enumerate}
\makeatletter
\newif\if@restonecol
\makeatother

\usepackage[ruled,linesnumbered,vlined]{algorithm2e}
\usepackage{algpseudocode}

\begin{document}
	\title{A unified framework of transformations based on the Jordan-Wigner transformation}

	\author{Qing-Song Li}
	\affiliation{CAS Key Laboratory of Quantum Information, University of Science and Technology of China, Hefei 230026, China}
	\affiliation{CAS Center for Excellence in Quantum Information and Quantum Physics, University of Science and Technology of China, Hefei 230026, China}
	\affiliation{Hefei National Laboratory, University of Science and Technology of China, Hefei 230088, China}
	\author{Huan-Yu Liu}
	\affiliation{CAS Key Laboratory of Quantum Information, University of Science and Technology of China, Hefei 230026, China}
	\affiliation{CAS Center for Excellence in Quantum Information and Quantum Physics, University of Science and Technology of China, Hefei 230026, China}
	\affiliation{Hefei National Laboratory, University of Science and Technology of China, Hefei 230088, China}
	
	\author{Qingchun Wang}
	\email{qingchun720@ustc.edu.cn}
	\affiliation{CAS Key Laboratory of Quantum Information, University of Science and Technology of China, Hefei 230026, China}
	\affiliation{CAS Center for Excellence in Quantum Information and Quantum Physics, University of Science and Technology of China, Hefei 230026, China}
	\affiliation{Hefei National Laboratory, University of Science and Technology of China, Hefei 230088, China}
	
	\author{Yu-Chun Wu}
	\email{wuyuchun@ustc.edu.cn}
	\affiliation{CAS Key Laboratory of Quantum Information, University of Science and Technology of China, Hefei 230026, China}
	\affiliation{CAS Center for Excellence in Quantum Information and Quantum Physics, University of Science and Technology of China, Hefei 230026, China}
	\affiliation{Hefei National Laboratory, University of Science and Technology of China, Hefei 230088, China}
	\affiliation{Institute of Artificial Intelligence, Hefei Comprehensive National Science Center, Hefei, Anhui, 230088, China}
	
	\author{Guo-Ping Guo}
	\affiliation{CAS Key Laboratory of Quantum Information, University of Science and Technology of China, Hefei 230026, China}
	\affiliation{CAS Center for Excellence in Quantum Information and Quantum Physics, University of Science and Technology of China, Hefei 230026, China}
	\affiliation{Hefei National Laboratory, University of Science and Technology of China, Hefei 230088, China}
	\affiliation{Institute of Artificial Intelligence, Hefei Comprehensive National Science Center, Hefei, Anhui, 230088, China}
	\affiliation{Origin Quantum Computing Hefei, Anhui 230026, China}
	\date{\today}
	\begin{abstract}
		Quantum simulation of chemical Hamiltonians enables the efficient calculation of chemical properties. Mapping is one of the essential steps in simulating fermionic systems on quantum computers. 
		In this work, a unified framework of transformations mapping fermionic systems to qubit systems is presented, and many existing transformations, such as Jordan-Wigner, Bravyi-Kitaev, and parity transformations, are included in this framework. 
		Based on this framework, the Multilayer Segmented Parity (MSP) transformation is proposed. The MSP transformation is a general mapping with an adjustable parameter vector, which can be viewed as a generalization of the above-mentioned mappings. Furthermore, the MSP transformation can adjust flexibly when dealing with different systems. 
		Applying these mappings to the electronic structure Hamiltonians of various molecules, the MSP transformation is found to perform better on the number of Pauli operators and gates needed in the circuit of Hamiltonian simulation. 
		The MSP transformation will reduce the qubit gate requirement for Hamiltonian simulation on noisy intermediate-scale quantum devices, and it will provide a much wider choice of mappings for researchers.
		
	\end{abstract}
	
	\maketitle
	\section{Introduction}
	The quantum computer (QC) was first driven forward by Feynman to simulate many-body quantum systems~\cite{feynmansimulating}, which is intractable for classical computers. Efficient simulation of many-body systems, especially fermionic systems, is essential, for it will lead to breakthroughs in quantum chemistry, materials, and other areas\cite{babbush2018low,reiher2017elucidating,cao2018potential,kassal2011simulating,lu2012quantum,aspuru2018matter}. 
	
	Although the realization of fault-tolerant quantum computing~\cite{fowler2012surface} has a long way to go, noisy intermediate-scale quantum (NISQ)~\cite{preskill2018quantum,bharti2022noisy} devices are promising to demonstrate quantum supremacy. Some potential applications have appeared, such as the variational quantum eigensolver (VQE)~\cite{peruzzo2014variational,mcclean2016theory}, which can be performed on NISQ devices to solve the electronic structure problems of molecules and materials. The quantum circuits should be as short as possible due to the high gate error rate and short dephasing time involved.
	
	As we know, electrons are fermions, which are antisymmetric indistinguishable particles. However, the qubits are distinguishable; so, the simulation of fermionic systems on quantum devices needs a mapping from fermion to qubit. Fermionic operators are mapped to multi-qubit Pauli operators, which are also referred to as Pauli strings. The length of Pauli strings, referred to as Pauli weight, determines the number of gates in the quantum circuit. Furthermore, it should be noted that two-qubit gates can only act on the nearest neighboring qubits in most quantum computation devices nowadays; thus, the Pauli strings should be contiguous, otherwise more gates are required to connect the non-nearest neighbor qubits. Thus, an excellent fermion-to-qubit mapping should improve the Pauli weight and the contiguity of Pauli strings.
	
	The most basic and widely used mapping is the Jordan-Wigner (JW) transformation~\cite{jordan1928pauli,somma2002simulating,nielsen2005fermionic,whitfield2011simulation}, and another primary mapping used is the parity transformation. The JW and the parity transformations map operators of $M$-orbital fermionic systems to Pauli strings of length $\order{M}$. 
	Combing the JW and parity transformations, Bravyi and Kitaev introduced the Bravyi-Kitaev (BK) transformation~\cite{bravyi2002fermionic,seeley2012bravyi,tranter2015b}, and then the BK-tree transformation~\cite{havlivcek2017operator} was introduced as a generalization of the BK transformation. The Pauli weight of the BK and BK-tree transformations is $\order{\log M}$. Although the Pauli weight is reduced from $\order{M}$ to $\order{\log M}$, the Pauli strings become discontiguous.  Those transformations have been compared in many works~\cite{tranter2018comparison,tranter2015b,seeley2012bravyi}. Besides the BK-tree transformation, the work~\cite{havlivcek2017operator} also introduced the Segmented Bravyi-Kitaev (SBK) transformation. The SBK transformation is defined on a  $W \times H$ lattice model, and the BK-tree transformation is applied to every row of the lattice. Similarly, the E-type auxiliary qubit mapping (E-type AQM)~\cite{steudtner2019quantum} applies the JW transformation to every row of the lattice and attaches one qubit to every row to store the parity of this row. 
	For a $W \times H$ lattice system, the Pauli weight of the SBK transformation is $\order{\log W +H} $, and the Pauli weight of the E-type AQM is $\order{W+H}$. The Pauli weight of the SBK transformation is usually greater than that of the BK and BK-tree transformations, but for nearest interaction models, it performs better. The Pauli weight of the E-type AQM is usually greater than that of the BK and BK-tree transformations. However, when we take the limitation of two-qubit gate manipulation into consideration, it is not the same case. We refer to those mentioned mappings as traditional mappings in the rest of the paper. They are all built based on the JW transformation.
	
	Another class of mapping schemes is based on the second mapping introduced by Bravyi and Kitaev, known as the Bravyi-Kitaev Superfast (BKSF) transformation\cite{bravyi2002fermionic,setia2018bravyi,chien2019analysis}. This mapping represents each fermion by a vertex on a graph, and the edges connecting two vertexes represent the interaction items of Hamiltonians. Then each edge is attached to a qubit. The Pauli weight of the BKSF transformation is $\order{d}$, where $d$ is the degree of the graph. The variants of the BKSF transformation are generalized superfast encodings (GSEs)~\cite{setia2019superfast} that require the same number of qubits as that of the BKSF transformation, but they have more favorable properties. Besides this, Verstraete and Cirac introduced the VC transformation~\cite{verstraete2005mapping}, which eliminates the string of Z operators by adding an auxiliary fermionic system.  More mappings~\cite{farrelly2014causal,whitfield2016local,steudtner2019quantum,derby2021compact,chen2022equivalence} were introduced using similar ideas as the VC transformation. The OpenFermion~\cite{mcclean2020openfermion} and ChemiQ~\cite{wang2021chemiq} packages have implemented some widely used transformations.
	
	Those mappings based on BKSF and VC transformations are local mappings, and they usually use many auxiliary qubits. It should be noted that the local mappings are usually designed for lattice systems and can't be used for molecular systems, while the traditional mappings are suitable for general systems and use no auxiliary qubits, except for E-type AQM. However, the traditional mappings are non-local. That is, the Pauli weight of traditional mappings increases with the size of systems. Compared with the JW and parity transformations, the BK and BK-tree transformations improve the Pauli weight from $\order{M}$ to $\order{\log M}$, but the Pauli strings are discontiguous. The E-type AQM is a compromise between the JW and BK transformations. Although it uses some auxiliary qubits, we will improve it.
	
	Our work concentrates on the traditional mappings. We present a framework of traditional mappings in which these mappings can be expressed in a unified form. What is more, a lot of potential transformations can be obtained more easily using this framework. Based on this framework, we propose the Multilayer Segmented Parity (MSP) transformation. The MSP transformation is a general transformation with a parameter vector, which can be adjusted according to the structure of fermion systems and quantum computation devices. The MSP transformation will reduce to the JW, BK-tree, and some other traditional mappings with corresponding parameter vectors. It also generates more useful transformations, such as an improved variant of the E-type AQM which uses no auxiliary qubits. We apply the MSP and some other traditional mappings to the electronic structure Hamiltonians of various molecules and then compare the numbers of Pauli operators and gates. The result is that the MSP transformation performs better than the other traditional mappings on almost all the molecules that we have tested. 
	
	The rest of this paper is organized as follows:
	In Sec.~\ref{background}, we will introduce the second quantization and the quantum circuit to simulate the evolution of Hamiltonians. In Sec.~\ref{unifiedframework}, a general framework is concluded to represent the traditional mappings. A mapping is defined by the summation sets, and they generate three other kinds of sets, which are convenient to represent the state and operator transformations. In Sec.~\ref{MSP}, the MSP transformation will be introduced based on our framework. In Sec.~\ref{Advantages}, we will discuss its advantages compared with other traditional mappings.
	In Sec.~\ref{perfermance}, we test the performance of the MSP and some other traditional mappings.
	The summary will be given in Sec.~\ref{conclusion}.

	\section{Background}\label{background}
	\subsection{Second quantization}\label{SecondQuantization}
	Due to the Pauli exclusion principle, each single-particle state contains at most one fermion. Therefore, each single-particle state is either occupied or empty, and usually a single-particle state is referred to as a spin-orbital. Consider an $M$-orbital fermionic system, whose state is in a $2^M$-dimensional Hilbert space spanned by basis states: $\{\ket{n_0,n_1,\dots,n_j,\dots,n_{M-1}}\}$, where $n_j\in \{0,1\}$ is the occupation number of the $j$th single-particle state. It is convenient to express the state and Hamiltonian of fermionic systems in terms of the creation operator $a_j^\dagger$ and the annihilation operator $a_j$, which are defined as follows,
	\begin{equation}
		\begin{split}\label{defing_fermionic_operators}
			a_j\ket{n_0,\dots,n_j,\dots,n_{M-1}}=\Gamma_j \delta_{1,n_j}\ket{n_0,\dots,0,\dots,n_{M-1}},\\
			a_j^\dagger \ket{n_0,\dots,n_j,\dots,n_{M-1}}=\Gamma_j \delta_{0,n_j}\ket{n_0,\dots,1,\dots,n_{M-1}} ,
		\end{split}
	\end{equation}
	where $\Gamma_j = (-1)^{\sum_{k=0}^{j-1}\hat{n}_k}$ is the phase factor, and $\hat{n}_j=a_j^\dagger a_j$ is the particle number operator. The creation and annihilation operators satisfy the following anticommutation relations:
	\begin{equation}
		\begin{split}
			&\{a_i,a_j\}=0,\\
			&\{a_i^\dagger,a_j^\dagger\}=0,\\
			&\{a_i,a_j^\dagger\}=\delta_{i,j}I,
		\end{split}
	\end{equation}
	where $\{A,B\}$ is defined as $\{A,B\}=AB+BA$. Such relations ensure antisymmetry of the wave function under the exchange of fermions.
	
	The Hamiltonian of interest, such as the electronic structure Hamiltonian of molecules, can be written in terms of creation and annihilation operators,
	\begin{equation}\label{Hamiltonian}
		H=\sum_{i,j}h_{ij}a_i^\dagger a_j+\frac{1}{2}\sum_{i,j,k,l} h_{ijkl}a_i^\dagger a_j^\dagger a_l a_k, 
	\end{equation}
	where the coefficients $h_{ij}$ and $h_{ijkl}$ are one- and two-electron integrals, respectively. The integrals can be calculated by the Psi4~\cite{parrish2017psi4}, PySCF~\cite{sun2018pyscf}, and ChemiQ~\cite{wang2021chemiq} packages. Other physical operators can also be written in terms of creation and annihilation operators, so we are mainly concerned about the transformation of fermionic states and creation (annihilation) operators. 
	
	\subsection{Hamiltonian simulation}\label{simulationOfHamiltonian}
\begin{figure}[h]
		\centering
		\includegraphics[width=0.48\textwidth]{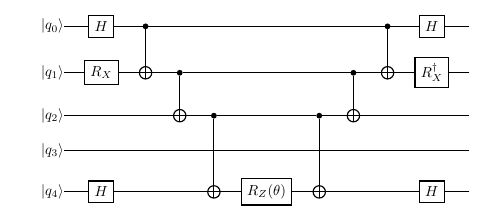} 
		\caption{The circuit to simulate the evolution of the one-item Hamiltonian $e^{-iH_1t/\hbar}$, where $H_1=h_1 X_0 Y_1 Z_2 X_4$. $R_X$ and $R_Z$ are rotation gates about the corresponding axes. The rotation angle $\theta$ is $2h_1t/\hbar$. }\label{examplecircuit}
	\end{figure}

	The mapped Hamiltonian can be written as the sum of Pauli strings,
	\begin{equation}\label{Paulistring}
		H_{\text{map}}=\sum_{l} h_lH_l,{} \; H_l= \prod_k P_k ,
	\end{equation}
	where $h_l$ is the coefficient and $P_k \in \{X_k,Y_k,Z_k\}$ is the one-qubit Pauli operator acting on the $k$th qubit. If there is only one item in the Hamiltonian, we can implement the evolution operator of the Hamiltonian $e^{-iHt/\hbar}$ directly, and the example circuit is depicted in Fig.~\ref{examplecircuit}~\cite{nielsen2002quantum}. The Hardmard and $R_X$ gates are applied to rotate the corresponding qubits into the X and Y basis, and CONT gates are used to compute the parity. Then the single-qubit gate, $R_Z(\theta)$, is applied to rotate the state. After this, more CNOT gates are used to un-compute the parity, and finally, Hardmard and $R_X^\dagger$ gates are applied to change the corresponding qubits back to the Z basis.

	One may want to simulate it sequentially when the Hamiltonian has more than one item. This is reasonable only in the condition that all items mutually commute. However, the items of a Hamiltonian usually do not commute with each other, and thus the Suzuki-Trotter approxmation~\cite{trotter1959product,suzuki1992general} is usually used. The first order of Suzuki-Trotter formulas is
	\begin{equation}\label{Suzuki}
		e^{(A+B)t}\approx (e^{At/n}e^{Bt/n})^n+\order{t^2/n},
	\end{equation}
	where $t$ is the evolution time, and $n$ is the number of simulation steps, which is named as the slice number. We can simulate the evolution of the Hamiltonian in a short time: $t/n$, and repeat it  $n$ times. 
	\begin{figure}[h]
		\centering

		\subfigure[]{
			\includegraphics[width=0.3\textwidth]{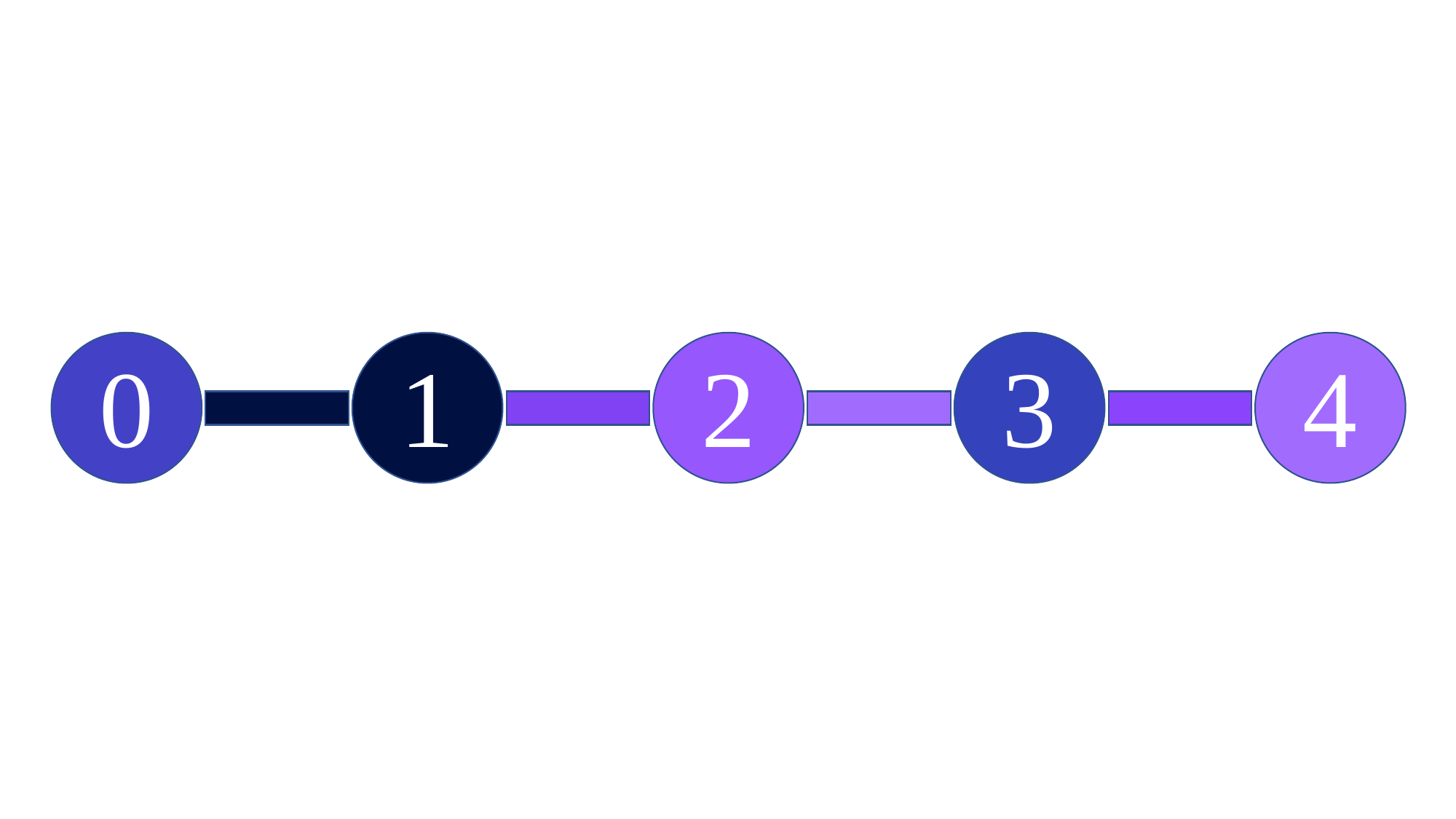}    }
		\subfigure[]{
			\includegraphics[width=0.45\textwidth]{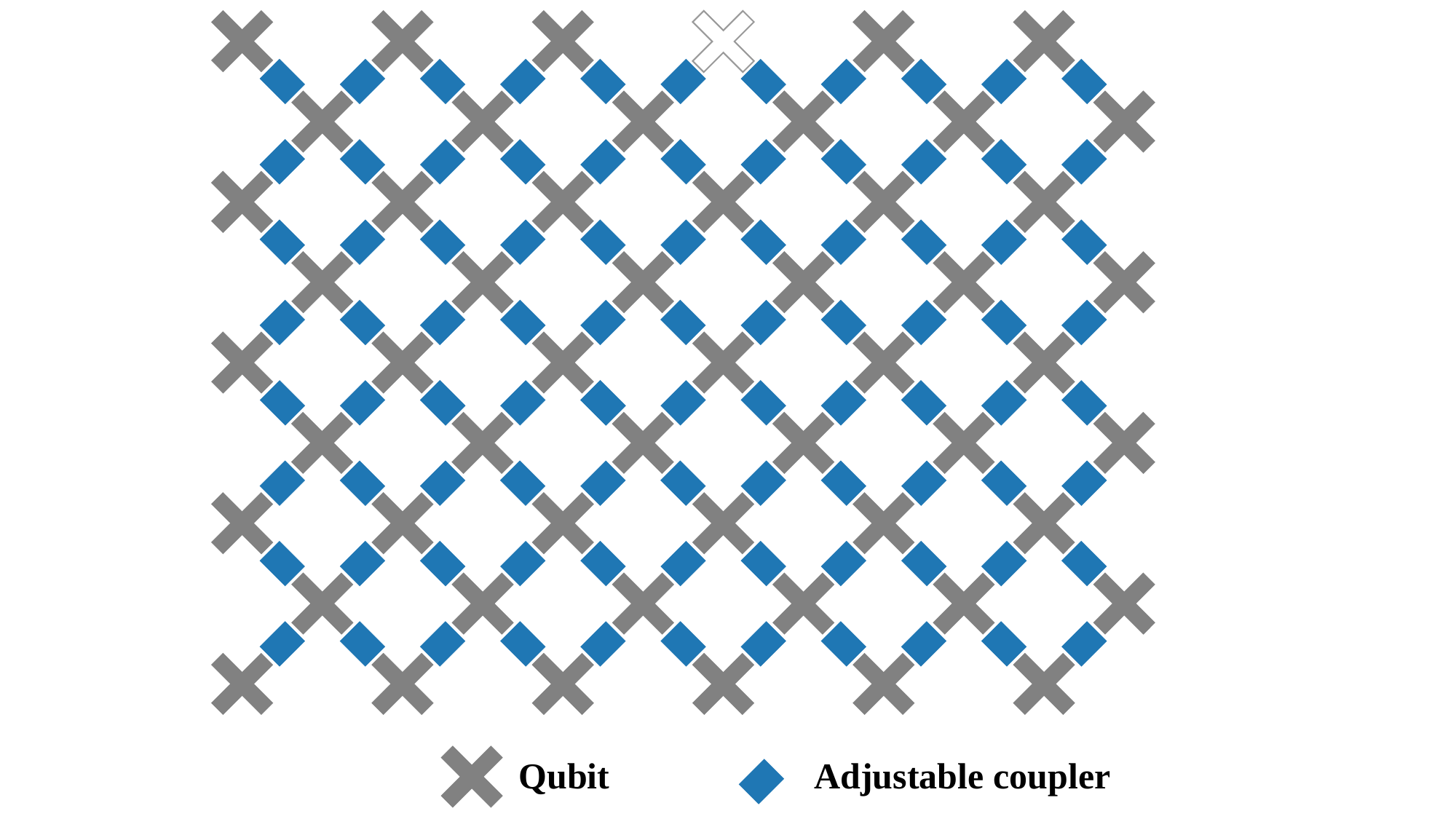}    }
		\caption{\label{chip} Layout of quantum processors.  a) IBM's Manila processor\cite{ibm_manila}. The circles represent qubits, and the thick lines are couplers connecting the nearest neighboring qubits. The qubit configuration of the Manila processor can be viewed as a one-dimensional lattice. b) Google's Sycamore processor\cite{arute2019quantum}. The white qubit in the first line is a corrupted qubit. The qubit configuration of the Sycamore processor can be viewed as a two-dimensional lattice. The couplers connect the nearest neighboring qubits. Two-qubit gates can only act on two qubits connected by a coupler.}	
	\end{figure}
	
	Here we can see that if the Pauli string is discontiguous, such as $X_0Y_1Z_2X_4$, it leads to CNOT gates (CNOT($q_2,q_4$)) acting on non-nearest neighboring qubits. However, if the two qubits are far from each other on the quantum device, it is difficult to implement the CNOT gate directly, and usually more SWAP gates are needed to swap the state on the target qubit to the qubit next to the control qubit. The number of SWAP gates is proportional to the distance between the two qubits on the quantum device.

	As shown in Fig.~\ref{chip}, the couplers only connect the nearest neighboring qubits. Thus, two-qubit gates cannot act on two non-nearest neighboring qubits directly. For instance, the CNOT(1,2) can be implemented on the Manila processor while the CNOT(1,3) can not act directly because there is no coupler connecting qubit 1 and qubit 3. 
	The qubit configurations of the two processors can be viewed as $n$-dimensional lattices. For the sake of discussion, we assume that qubit configurations of quantum devices are $n$-dimensional lattices in this paper.

	\section{Unified framework of traditional transformations}\label{unifiedframework}
	In our framework, the transformations of states are represented by summation sets and the transformations of creation (annihilation) operators are represented by the parity, flip, and update sets.
	
	\subsection{Representation of the occupation state}
	The basis state transformation is described by summation sets:
	\begin{equation}\label{statetrans}
		\begin{split}
			\ket{n_0,n_1,\dots,n_{M-1}} \rightarrow \ket{x_0,x_1,\dots, x_{M-1}},\\
			x_j=n_j+\sum_{k \in S(j)}n_k \pmod 2,
		\end{split}
	\end{equation}
	where $S(j)$ is the summation set whose elements are some sites with indices less than $j$. Note that all the additions are binary additions in this paper. In the parity transformation, the $S(j)$ is $\{0,1,2,\dots,j-1\}$, and in the JW transformation, the $S(j)$ is an empty set. A traditional mapping completely depends on the summation sets.
	
	One can also use the transformation matrix $A$ to represent the state transformation. The matrix $A$ and $S(j)$ sets can be generated mutually:
	\begin{equation}
		A_{jk}=\left\{
		\begin{array}{ll}
			1, & \text{if $k \in S(j)$ or $k=j$}\\
			0, & \text{otherwise.}
		\end{array}
		\right. ,
	\end{equation}
	\begin{equation}
		S(j)=\{k|\text{$k < j$ and $A_{jk}=1$}\},
	\end{equation}
	where $A_{jk}$ is the entry of matrix $A$. Note that the indices of the sites in $S(j)$ are smaller than $j$ in our definition.
	
	To simplify the discussion of the operator transformation, two constraints are imposed on summation sets:
	\begin{enumerate}[i]
		\item For any $i<j<k\in \{0,1,\dots,M-1\}$, if $i \in S(j)$ and $j \in S(k)$, then $i \in S(k)$.
		\item For any $i<j<k\in \{0,1,\dots,M-1\}$, if $i \in S(j)$ and $j \notin S(k)$, then $i \notin S(k)$.
	\end{enumerate}
	
	One can verify that the summation sets of the traditional mappings satisfy those two constraints. Actually, the two constraints are not necessary for more general mapping schemes, but with those two constraints, we can generate three other kinds of sets uniquely and build a tree structure. 
	
	We give an example of a seven-femion transformation. The summation sets are built almost randomly but satisfy the two conditions,
	\begin{equation}
		\left\{
		\begin{array}{l}
			S(0)=\varnothing\\
			S(1)=\varnothing\\
			S(2)=\{1\}\\
			S(3)=\varnothing \\
			S(4)=\{0,3\}\\
			S(5)=\{0,3,4\}\\
			S(6)=\{0,1,2,3,4,5\}
		\end{array}
		\right..
	\end{equation}

	The summation set of the last site, $S(6)$, contains all the sites with indices less than $6$, which is the same with most of the traditional mappings, but it is not always the case. 
	Then we write the state transformation according to the summation sets in Eq.~(\ref{statetrans}):
	\begin{equation}
		\left\{
		\begin{array}{l}
			x_0=n_0\\
			x_1=n_1\\
			x_2=n_2+n_1\\ 
			x_3=n_3\\
			x_4=n_4+n_0+n_3\\
			x_5=n_5+n_0+n_3+n_4\\
			x_6=n_6+n_0+n_1+n_2+n_3+n_4+n_5
		\end{array}
		\right..
	\end{equation}
	
	\subsection{The filp, parity, and update sets}
	In the last example, we can also rewrite $x_j$ as the sum of $n_j$ and some $x_i$:
	\begin{equation} \label{filpEquation}
		\left\{
		\begin{array}{l}
			x_0=n_0\\ 
			x_1=n_1\\
			x_2=n_2+x_1\\
			x_3=n_3\\
			x_4=n_4+x_0+x_3\\
			x_5=n_5+x_4\\ 
			x_6=n_6+x_2+x_5
		\end{array}
		\right..
	\end{equation} 
	Actually, for the traditional mappings, $x_j$ can be uniquely written in terms of $n_j$ and some $x_k~(k<j)$,
	\begin{equation}\label{StateTransFset}
		x_j = n_j+\sum_{k \in F(j)} x_k,
	\end{equation}
	where $F(j)$ is the flip set defined as:
	\begin{equation} \label{defFlipSets}
		F(j)=\{k|k \in S(j)\text{ and } k \notin S(i) \text{ for any } i<j \}.
	\end{equation}
	
	Before discussing the operator transformation, it's necessary to introduce two other kinds of sets: the parity set $P(j)$ and the update set $U(j)$. The two kinds of sets are defined as:
	\begin{equation}
		P(j)=\{k|k<j \text{ and }k \notin S(i) \text{ for any } i<j \},
	\end{equation}
	\begin{equation}
		U(j)=\{k|j \in S(k) \}.
	\end{equation}
	According to the definition, one can find that $F(j)$ is the subset of $P(j)$. $F(j)$ and $P(j)$ have no common elements with $U(j)$.
	$U(j)$ is used to update the states of some qubits, which change with the occupation number of the $j$th site. $P(j)$ is applied to obtain the parity $p(j)$:
	\begin{equation}\label{paritycount}
		p(j)=\sum_{i=0}^{j-1} n_i=\sum_{k \in P(j)} x_k.
	\end{equation}
	
\begin{figure}[h]
		\centering
		\includegraphics[width=0.2\textwidth]{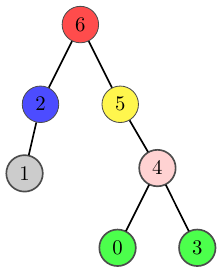} 
		\caption{A seven-fermion mapping tree. The child node of node 5 is node 4; thus, $F(5)$ is $\{4\}$ . The descent nodes of node 5 are node 0 and node 3; so, $S(5)$ is $\{0, 3, 4\}$. The ancestor node of node 5 is node 6; thus, $U(5)$ is $\{6\}$.   Node 2 and node 4 have no father nodes with indices smaller than $5$; so, $P(5)$ is $\{2,4\}$.  }\label{7-fermion tree}
	\end{figure}

	To show the relations of the four kinds of sets more clearly, we construct a mapping tree via the flip sets as shown in Fig.~\ref{7-fermion tree}. In the mapping tree, node $j$ corresponds to the $j$th site. The elements of $F(j)$ are the child nodes of node $j$, and the elements of $S(j)$ are the descent nodes of node $j$. The elements of $U(j)$ are the ancestor nodes of node $j$. The elements of $P(j)$ are the root nodes with indices less than $j$; in other words, they are the nodes that have no father nodes with indices smaller than $j$. Given a mapping tree, we can also generate these sets easily.
	
	\subsection{Representation of creation and annihilation operators}
	Recall that $a_j~(a_j^\dagger$) translates the jth occupation number from $1~(0)$ to $0~(1)$ and from $0~(1)$ to nothing. In the JW transformation, a pair of one-qubit operators is used to represent this,
	\begin{equation}\label{QoperatorDef}
		Q_j^\dagger \ket{0}_j=\ket{1}_j,~ Q_j^\dagger \ket{1}_j=0,~ Q_j\ket{1}_j=\ket{0}_j,~ Q_j\ket{0}_j=0,
	\end{equation}
	where $Q_j^\dagger$ and $Q_j$ are one-qubit creation and annihilation operators. And they can be written in terms of Pauli operators,
	\begin{equation}
		Q_j=\frac{X_j+iY_j}{2}, ~ Q_j^\dagger=\frac{X_j-iY_j}{2},
	\end{equation}
	where $X_j$ and $Y_j$ are Pauli X and Y operators acting on the $j$th qubit.
	However, for other traditional transformations, according to Eq.~($\ref{StateTransFset}$), $x_j$ will flip compared with $n_j$ if $\sum_{k \in F(j)} x_k =1$, thus $Q_j$ and $Q_j^\dagger$ should be replaced by:
	\begin{equation}
		\begin{split}
			&Q_j \rightarrow \frac{(I+Z_{F(j)})}{2}Q_j + \frac{(I-Z_{F(j)})}{2}Q_j^\dagger,\\
			&Q_j^\dagger \rightarrow \frac{(I+Z_{F(j)})}{2}Q_j^\dagger + \frac{(I-Z_{F(j)})}{2}Q_j,
		\end{split}
	\end{equation}
	where $Z_{F(j)}$ is a multi-qubit operator applying Pauli Z operators to the qubits in $F(j)$. $\frac{(I+Z_{F(j)})}{2}$ and $ \frac{(I-Z_{F(j)})}{2}$ are projection operators to check whether $\sum_{k \in F(j)} x_k $ is 0 or 1. 
	
	The phase factor in Eq.~(\ref{defing_fermionic_operators}) is obtained using Eq.~(\ref{paritycount}),
	\begin{equation}
		\Gamma_j=(-1)^{\sum_{i=0}^{j-1}\hat{n}_{i}} \rightarrow Z_{P(j)},
	\end{equation}
	where $Z_{P(j)}$ is a multi-qubit operator acting Pauli Z operators on the qubits in $P(j)$. Moreover, if $j \in S(k)$, the $k$th qubit will change from 1 (0) to 0 (1) when the occupation number of the $j$th site changes; thus, it's necessary to apply $X_{U(j)}$ to update these qubits in $U(j)$, where $X_{U(j)}$ is a multi-qubit operator applying Pauli $X$ operators to the qubits in $U(j)$. 
	Based on those sets, the representation of the operator transformation is concise,
	\begin{equation}\label{operatortrans}
		\begin{split}
			a_j&=Z_{P(j)}[(\frac{1+Z_{F(j)}}{2})Q_j+(\frac{1-Z_{F(j)}}{2})Q_j^\dagger]X_{U(j)}\\
			&=Z_{P(j)}(\frac{X_j+iZ_{F(j)}Y_j}{2})X_{U(j)}\\
			&=(\frac{Z_{P(j)}X_j+iZ_{P(j)/F(j)}Y_j}{2})X_{U(j)},\\
			a_j^\dagger&=(\frac{Z_{P(j)}X_j-iZ_{P(j)/F(j)}Y_j}{2})X_{U(j)},
		\end{split}
	\end{equation}
	where $P(j)/F(j)$ is the set that consists of all elements in $P(j)$ while not in $F(j)$.
	
	In the JW transformation, $S(j)$, $F(j)$, and $U(j)$ are empty sets and $P(j)$ contains all sites with indices less than $j$. In the parity transformation, the $S(j)$ contains all sites with indices smaller than $j$, whereas $P(j)$ and $F(j)$ only contain one site, i.e., $j-1$ $(j >0)$, and $U(j)$ contains all sites with indices greater than $j$. Ref~\cite{bravyi2002fermionic} gives the expressions of the four kinds of sets for the BK transformation when $M$ is the power of 2. The BK-tree transformation is suitable for any $M$, but there are no brief math expressions of these sets. Nevertheless, they give an algorithm to build the Fenwick tree, and the four kinds of sets can be generated via the Fenwick tree. We give an algorithm in Appendix \ref{appendix AlgOfThreeSets} to generate these sets in Eq.~(\ref{operatortrans}), and it is suitable for all the traditional mappings. To design a new mapping, one only needs to build $S(j)$ sets, and then the mapping is realized by Algorithm. \ref{algorithm GeneratingSets}. 
	
	The four kinds of sets give physical intuition of the representation of mappings, and they all have clear physical meanings. Using Eq.~(\ref{operatortrans}), the traditional mappings can be represented as this unified form. Although the representation of the JW and parity transformations is simple, it is hard to express the BK (BK-tree) and MSP transformations explicitly because the results of different creation and annihilation operators are quite different. For an example of a 12-fermion system, the result of $a_3$ is $ Z_2 \frac{X_3+Y_3}{2} X_4 X_5 X_{11}$ mapped by the BK-tree transformation, while the result of $a_8$ is $Z_5 \frac{Y_8+Z_7 X_8}{2}{X_{11}}$. The indices of Pauli strings are quite different, but they have a unified form.
	
	According to Eq.~(\ref{operatortrans}), one can find that the asymptotic scaling of the Pauli weight of mappings depends on the orders of $P(j)$ and $U(j)$, where the order of a set is the number of elements of the set. The JW and parity transformations do not balance the two sets well, for the order of $P(j)$ or $U(j)$ is too large. The BK and BK-tree transformations are compromises between the JW and parity transformations, where both the orders of $P(j)$ and $U(j)$ are not too large. Usually, making $P(i)$ smaller will pay the price for the larger order of $U(j)$. Thus, a good mapping should balance the two sets well. 
	
	However, the Pauli weight is not the only criterion to judge mappings. The contiguity is also crucial when implementing the evolution operator of Hamiltonians, which is a key procedure in some quantum algorithms~\cite{kitaev1995quantum,abrams1999quantum}. However, in general, improving the contiguity will increase the Pauli weight. Therefore, we must weigh them when choosing or designing a mapping for a system. We will give further discussion later.
	\section{Multilayer Segmented Parity Transformation}\label{MSP}
	
	\subsection{Introduction of the MSP transformation}
	
		\begin{figure}[h]
		\centering

		\subfigure[]{\label{qd_1d}
			\includegraphics[width=0.45\textwidth]{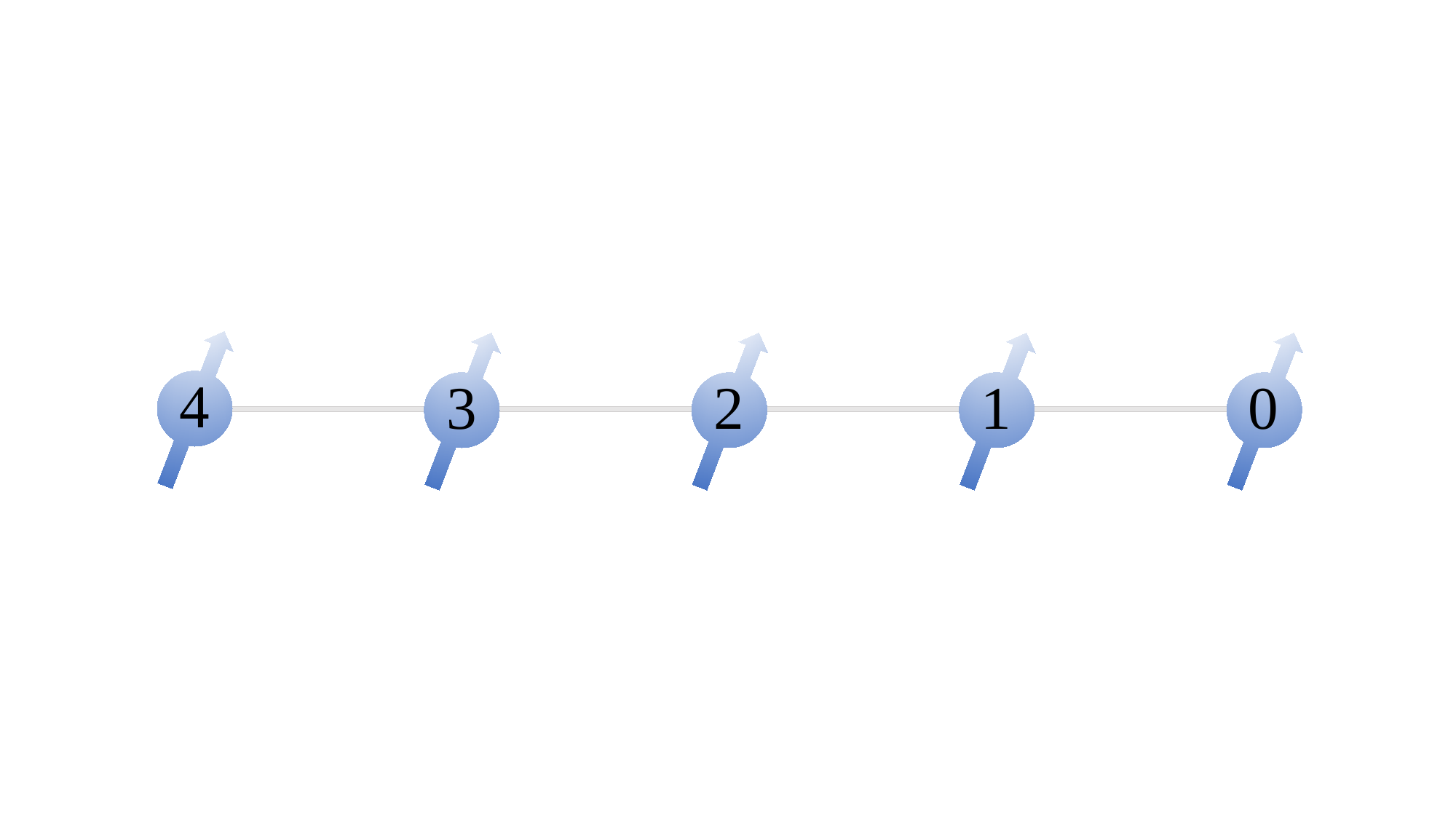}    }
		\subfigure[]{\label{qd_2d}
			\includegraphics[width=0.45\textwidth]{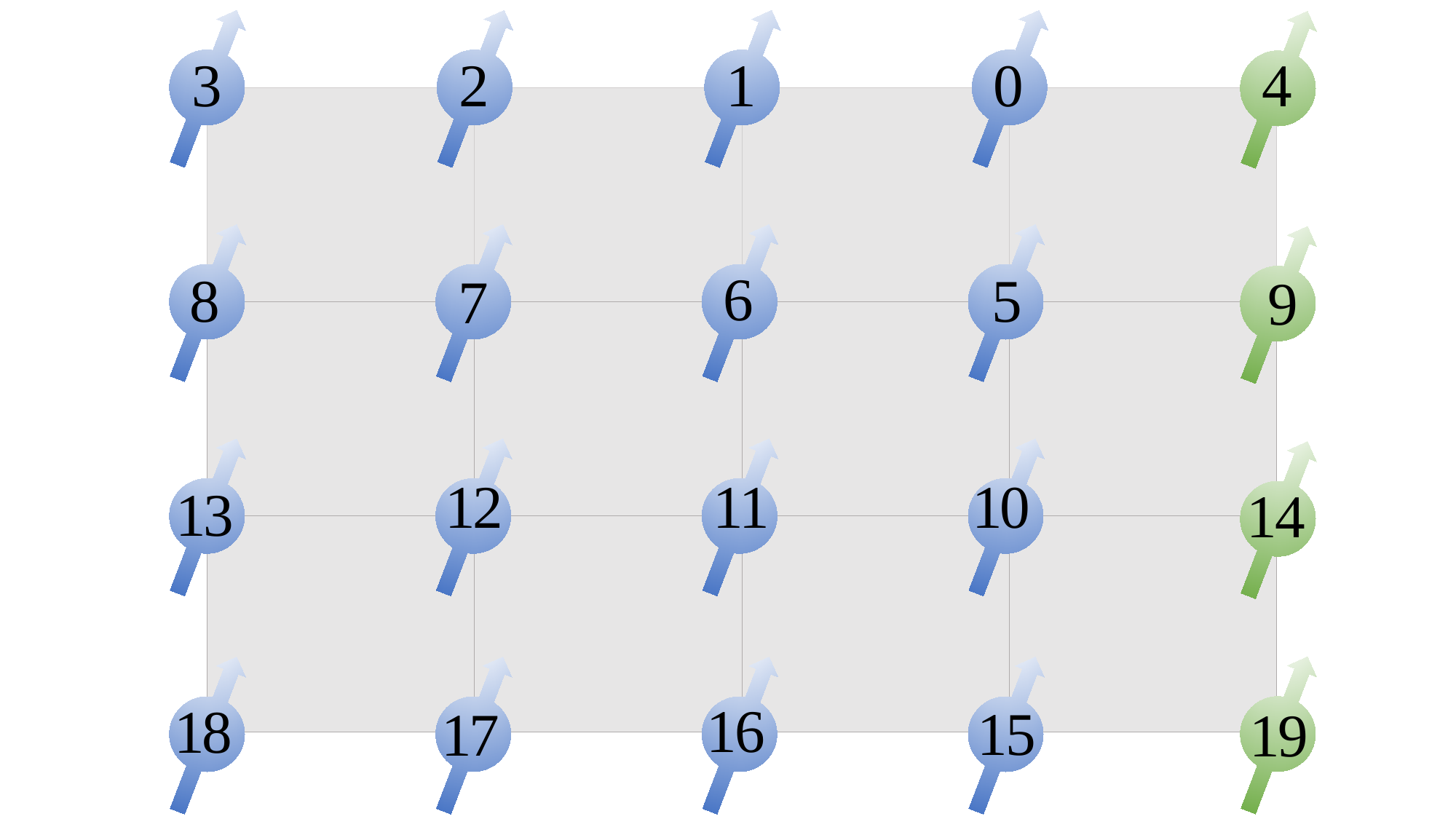}    }
		\subfigure[]{\label{qd_3d}
			\includegraphics[width=0.45\textwidth]{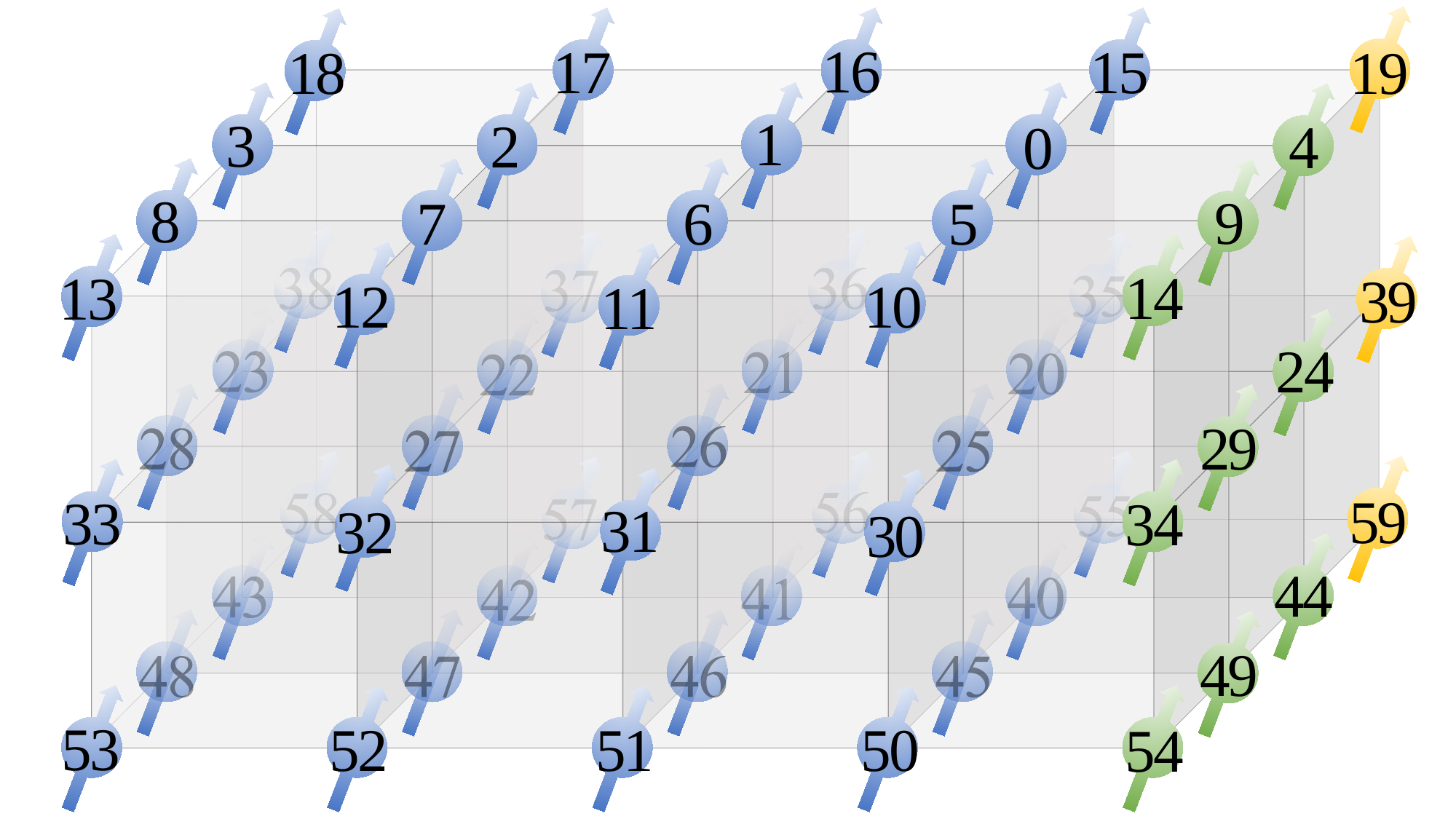}    }
		\caption{\label{qd} Qubit configurations of quantum processors.  a) A one-dimensional quantum device. b) A two-dimensional quantum device. c) A three-dimensional quantum device. The JW Pauli strings are contiguous on any dimensional quantum device, whereas the BK Pauli strings are not contiguous on these quantum devices in general. The Pauli strings of an $L$-layer mapping cannot be contiguous on a quantum device with a dimension lower than $L$. }	
	\end{figure}

	Before introducing the MSP transformation, we first show how the mapping is constructed. The idea of constructing it is from keeping Pauli strings contiguous on different quantum devices. The JW Pauli strings are contiguous on $n$-dimensional quantum devices. However, the Pauli weight of the JW transformation is very high. We want to reduce the Pauli weight and keep the Pauli strings contiguous. Fortunately, our wish is satisfied on two-dimensional quantum devices. As shown in Fig. \ref{qd_2d}, we let the last qubit (green) of each line record the total number of fermions (parity) in this line and the other qubits (blue) record their occupation numbers. Then the Pauli weight reduces to $\order{M^{1/2}}$ and the Pauli strings are still contiguous. We find that a further reduction of Pauli weight is available on three-dimensional quantum devices, as shown in Fig. \ref{qd_3d}. The last qubit (yellow) of each plane records the parity of this plane and the other qubits are the same with the two-dimensional case. The Pauli weight is $\order{M^{1/3}}$. In general, the Pauli weight can be reduced to $\order{M^{1/D}}$ on a $D$-dimensional quantum device. (Although one- and two-dimensional quantum devices are the most popular, it is possible to build higher-dimensional quantum devices. For instance, we can build a three-dimensional quantum device on a two-dimensional quantum chip.) Interestingly, the BK transformation is just the special case on the $\log_2 M$-dimensional quantum device of size $2 \times 2 \cdots \times 2$.
	One can notice that the proper mappings are quite different on different quantum devices. Thus, it is necessary to introduce a parameter vector to denote these mappings. Then, we can define these mappings in a unified form, the Multilayer Segmented Parity (MSP) transformation.

	In the MSP transformation, the sites (orbitals or fermions) will be divided into many segments. Suppose the number of sites, $M$, can be factorized as $M=v_1 \times v_2 \times \dots \times v_L$, where $L$ is the layer of MSP transformation, and we can  use a row vector to denote this: $\vec{V}=(v_1, v_2, \dots, v_L)$. Note that $v_j$ is not necessarily a prime number. Then the $M$ sites will be divided into segments according to $\vec{V}$ as follows:
	\begin{enumerate}[step 1]
		\item   The $M$ sites are equally divided into $v_1$ segments, and we define the last site of each segment as the parity site of the segment. 
		\item Every segment is equally divided into $v_2$ subsegments and the last site of each subsegment is defined as the parity site of the corresponding subsegment.
		\item Repeat the division $L$ times.
	\end{enumerate}
	
	\begin{figure}[h]
		\includegraphics[width=0.48\textwidth]{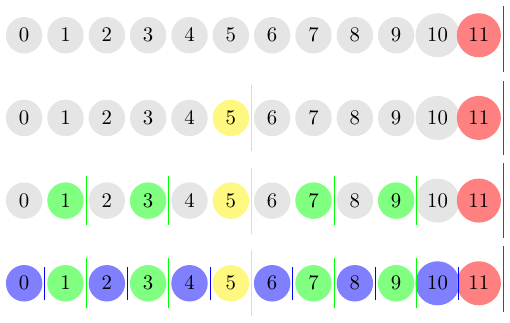} 
		\caption{The procedure of the division for the 12-fermion transformation with $\vec{V}=(1,2,3,2)$. The red line is the first layer division line, and the red site is the parity site of the whole, so $S(11)$ is $\{0,1,\dots,9 ,10\}$. The yellow line is the second layer division line, and the yellow site is the parity site of the corresponding segment; thus, $S(5)$ is $\{0,1,2,3,4\}$. The green lines are the third layer division lines, and the green sites are the parity sites of the corresponding subsegments; so, $S(1)$ = $\{0\}$, $S(3)$ = $\{2\}$, $S(7) = \{6\}$ and $S(9) = \{8\}$. The blue lines are the fourth layer division lines, and the blue sites are the parity sites of the corresponding smallest segments; thus, $S(0) = S(2) =S(4)= S(6) = S(8) = S(10) = \varnothing$. Every S(j) is defined after all divisions.}\label{12-fermion MSPT}
	\end{figure}
	Note that the segment and the subsegment are relative concepts, so if not specifically stated, we refer to a subsegment as a segment too. One can notice that there is conflict during the steps. The parity qubit of a segment may be again defined as the parity qubit of smaller segments in the following steps. To eliminate this contradiction, we appoint that a qubit is the parity qubit of the segment when it is defined for the first time. To illustrate the procedure more clearly, we give an example of the 12-fermion MSP transformation with $\vec{V}=(1,2,3,2)$, and the progress is shown in Fig.~\ref{12-fermion MSPT}. We can see in step 2 that node 5 is defined as the parity qubit of the segment $(0,1,2,3,4,5)$, while in step 3, node 5 is the last node of the segment $(4,5)$. If node 5 is defined as the parity qubit of the segment $(4,5)$, it will conflict with step 2, so we should keep the former definition. 
	
\begin{figure}[h]
		\includegraphics[width=0.4\textwidth]{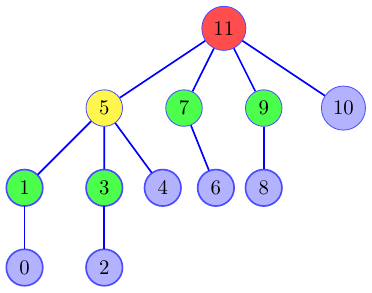} 
		\caption{The mapping tree of the 12-fermion MSP transformation with $\vec{V}=(1,2, 3, 2)$.}\label{12-fermion tree}
	\end{figure}

	The MSP transformation uses the qubit $q_j$ to store the parity of a segment if site $j$ is the parity site of this segment and to store the occupation number of the $j$th site if it is not the parity site of any segment. Actually, in the MSP transformation, every site has been defined as a parity site. As discussed in Sec.~\ref{unifiedframework}, we can also build a mapping tree of the MSP transformation. The mapping tree of the 12-fermion MSP transformation with $\vec{V}=(1,2,3,2)$ is shown in Fig.~\ref{12-fermion tree}. According to $S(j)$ or the mapping tree, it is easy to get three other kinds of sets, and then the representation of the MSP transformation is obtained.
	
	\subsection{The Pauli weight of the MSP transformation}
	The Pauli weight of the MSP transformation is $(\sum_{l=1}^{L} (v_l-1))+1$, and the proof is given in Appendix \ref{Pauli weight of MSP}. We consider the special case that all entries of $\vec{V}$ are the same:$v_1=\dots=v_L=v$, and $L=\log_v M$. The Pauli weight of this case is $(v-1)\log_v M+1$, and it grows with $v(v\geq2)$. Thus, the Pauli weight reaches its minimum value at $v=2$, which is $\log_2 M+1$, the Pauli weight of the BK transformation. In other words, the BK-tree transformation is the best case of MSP transformation when only considering the Pauli weight and $M$ is large enough. However, when considering the number of Pauli operators of a Hamiltonian, the BK transformation is not always the best choice because the number of Pauli operators depends on the form of specific Hamiltonians. In practice, when applying the MSP transformation with a proper vector to the electronic structure Hamiltonian of some molecules, the total number of Pauli operators is less than that of the BK transformation. And it should be noted that the Pauli weight is not the only metric to determine which mapping is the most suitable. The contiguity of Pauli strings and other metrics are also important.
	
	\subsection{Generlization of the MSP transformation}
	The original MSP transformation is only suitable for the case that $M$ can be factorized into some factors. If $M$ is a prime number or can't be factorized into small factors, the choice of the transformation parameter vector is restricted to a large extent. We will generalize the MSP transformation and make it suitable for any $M$. In the generalized MSP transformation, the transformation parameter vector is adjustable.
	
	Suppose a segment has $M_l$ sites (fermions), and we need to divide them into $v_l$ subsegments, but it is a problem if $M_l$ is not a multiple of $v_l$. It cannot be divided equally in this case, but we may divide it unequally. Suppose $M_l=u*v_l+r$ and $0 \leq r \leq v_l$, and we divide $u+1$ fermions into each subsegment for the front $r$ subsegments and $u$ fermions into each subsegment for the rest of the subsegments. In this division strategy, there is no need to impose the condition that $v_l$ is a factor of $M$, and $v_l$ can be any integer$(\geq 2)$, even greater than $M_l$. The generalized MSP transformation contains the case of the original one; hence, if not otherwise specified, the MSP transformation means the generalized one in the following. The MSP transformation reduces to the BK-tree transformation when we let $v_1=\dots =v_L=2$. 
	
	It is not easy to give the expressions of the four kinds of sets of the generalized MSP transformation, but we give the algorithm in Appendix \ref{appendix alg} to generate the summation sets. $S(j)$ is built by the function \textbf{Segment}($L,R,k)$, where $L$(left) is the first site and $R$(right) is the last site of the segment, and $k$ is used to divide the segment into $v_k$ subsegments. In essence, the \textbf{Segment}($L,R,k)$ is the generalization of \textbf{Fenwick($L,R$)}~\cite{havlivcek2017operator}. Then we invoke Algorithm \ref{algorithm GeneratingSets} to generate $U(j)$, $P(j)$ and $F(j)$ sets.

	\section{Advantages of the MSP transformation}\label{Advantages}
	\subsection{A general transformation}
	The MSP transformation is a general transformation that contains most of the traditional mappings. For example, the MSP transformation reduces to the JW transformation when $\vec{V}=(M)$. If we set the parameter $\vec{V}$ as $(1,2,2,\dots,2)$, the MSP transformation becomes the BK transformation when $M$ is the power of 2, and it becomes the BK-tree transformation when $M$ is not a power of 2. Moreover, it becomes the SBK transformation when setting $\vec{V} = (H,1,2,\dots,2)$.
	
	In particular, when setting $\vec{V}$ as $(H,W)$, we get a very useful version of the MSP transformation. Its total segment layer $L$ is 2, so we name it the 2SP transformation. The 2SP transformation is an improvement of the E-type AQM, for the 2SP transformation uses no auxiliary qubits and the Pauli weight of the 2SP transformation is no greater than that of the E-type AQM. Although the Pauli weight of the 2SP transformation is $\order{\sqrt{M}}$ when $H =W = \sqrt{M}$, which is greater than $\order{\log M}$, the Pauli weight of the BK transformation. However, when considering the limitations of quantum devices, the 2SP transformation is even better than the BK transformation when comparing the number of actually used gates. Table \ref{compMSP} shows the $\vec{V}$ of the MSP transformation corresponding to the traditional mappings.

	\begin{table}[!htbp]
		\centering
		\caption{
			The parameter vectors of the MSP transformation corresponding to the traditional mappings.  \label{compMSP}}
		\begin{tabular}[c]{ll}
			\hline \hline
			Mapping~	&Parameter vector \\
			\hline
			JW			& $(M)$	\\
			Parity			&$(1,1,\dots,1)$ \\
			2SP (E-type AQM)	& $(H,W)$ \\
			SBK			& $(H,1,2,2,\dots,2)$	\\
			BK (BK-tree)			& $(1,2,2,\dots,2)$	\\	
			\hline \hline
		\end{tabular}
	\end{table}
	Besides these mappings, the MSP transformation also generates many other mappings. For instance, when setting $\vec{V}=(1,M)$, we get a variant of the JW transformation. Different mappings are suitable for different systems, and we will discuss this in the following section.

	\subsection{Adjustable}
	The MSP transformation is adjustable so that we can choose the proper parameter vector when facing different systems. For molecular systems, the JW and BK transformations are widely used, but we can adjust $\vec{V}$ according to the total number of qubits and the structures of molecules. With proper vectors, the numbers of Pauli operators and gates of the MSP transformation can always be made less than those of the JW and BK transformations. Moreover, if the system is particle number conserved, the qubit reduction can be used. For instance, in the electronic structure problem of molecules (non-relativistic), the total number of electrons and total $s_z$ value are conserved, so we can rearrange the orbitals, making spin-up orbitals in the front half and spin-down orbitals in the back half. When applying the MSP transformation with $v_1=2$ to this system, the qubit $q_{M/2-1}$ stores the number of spin-up electrons and the qubit $q_{M-1}$ stores the number of spin-down electrons. Thus, the states of the two qubits remain unchanged when simulating this system, and then two qubits can be saved. 
	
	The lattice models are important in physics, such as the fermion Hubbard model~\cite{hubbard1963electron}. The main feature of the model is that the interaction items appear only between nearest-neighboring sites. For a one-dimensional lattice with $M$ sites, the JW transformation is proper, which is the same as the MSP transformation with $\vec{V}=(M)$. And we can also apply the MSP transformation with $\vec{V}=(1,M)$. In this transformation, the front $M-1$ qubits still store the corresponding occupation number, while the last qubit stores the particle number of all sites. It can save a qubit if the system is particle number conserved.
	
	For two-dimensional lattices, such as a lattice of size: $H\times W$ , we can apply the MSP transformation with $\vec{V}=(H,1,2,\dots,2)$, which is the same as the SBK transformation. The Pauli weight of the SBK transformation is $\order{H+\log W}$, which is greater than the Pauli weight $\order{\log M}~(M=W \times H)$ of the BK transformation for general systems. But the Pauli weight of the SBK transformation reduces to $\order{\log W}$ when the interaction is nearest-neighboring. 
	
	As for a three-dimensional lattice of size: $L\times W \times H~(L<W<H)$, the MSP transformation with $\vec{V}=(H,1,2,\dots,2)$ is suitable. If we restrict interaction nearest-neighboring, the Pauli weight of this transformation will reduce to $\order{\log (WL)}$. 
	\subsection{Considering quantum devices}\label{device}
	
	\begin{figure}[h]
		\centering
		
		\subfigure[]{
			\includegraphics[width=0.19\textwidth]{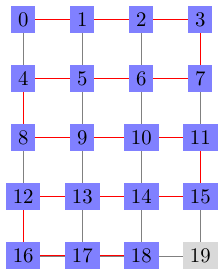}    }
		\subfigure[]{
			\includegraphics[width=0.19\textwidth]{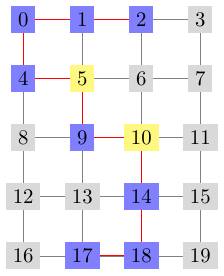}    }\\
		\subfigure[]{
			\includegraphics[width=0.21\textwidth]{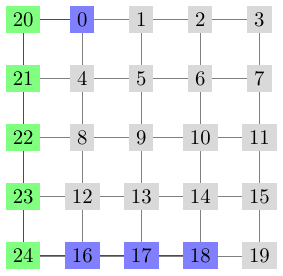}    }
		\subfigure[]{
			\includegraphics[width=0.18\textwidth]{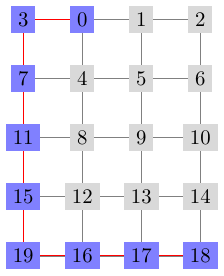}    }
		
		\caption{\label{actqubit} Qubits and two-qubit gates used to implement the evolution of an item of a Hamiltonian, $a_0a_{18}^\dagger + a_{18}a_0^\dagger$.  a) JW, b) BK, c) E-type, d) 2SP. The Hamiltonian is a 20-fermion Hamiltonian of molecular systems and the structure of the quantum device is a $5 \times 4$ lattice. The nodes represent qubits. The grey nodes are the qubits on which no operator acts. The blue nodes are qubits on which operators act. The green nodes are auxiliary qubits. The yellow nodes are qubits on which no operator acts but are used to connect the non-nearest neighboring qubits. The red lines represent two-qubit gates acting on the corresponding qubits. }	
	\end{figure}

	Here we take the connection limitations of quantum devices into consideration. Suppose $w$ is the square root of $M$, which is the proper setting to balance the Pauli weight and the contiguity of Pauli strings, then the Pauli weight of the 2SP transformation is $\order{\sqrt{M}}$, which is greater than the Pauli weight of the BK transformation but much smaller than that of the JW transformation when $M$ is large enough. Comparing the BK and JW transformations, the Pauli strings of the JW transformation are too long and the Pauli strings of the BK transformation are too discontiguous. The 2SP transformation combines the advantages of the JW and BK transformations and overcomes their disadvantages to some degree. The Pauli weight of the 2SP transformation is reduced a lot compared with the JW transformation, and the Pauli strings of the 2SP transformation are contiguous on 2D lattice quantum devices. 
	
	We give an example of the 20-fermion system. Considering an item of a Hamiltonian, $a_0 a_{18}^\dagger +a_{18}a_0^\dagger$, it is mapped by different mappings, and the qubits on which Pauli operators act and the two-qubit gate used are shown in Fig. \ref{actqubit}. The numbers of Pauli operators and gates for the JW transformation are much greater than those of the three other mappings. The number of Pauli operators for the BK transformation is eight, less than that of the E-type AQM and the 2SP transformation. However, it uses more two-qubit gates than the two other mappings. The defect of discontiguous Pauli strings of the BK transformation will become more serious when the size of systems goes up.

	The MSP transformation and its variants contain many traditional transformations, such as JW, BK and SBK transformations. Thus one can just use the MSP transformation instead of other traditional mappings in practice. Before choosing which mapping to use, we should consider the benefits and costs of these mappings carefully. The benefits and costs depend on the structures of Hamiltonians and quantum devices. We just give a brief discussion of choosing proper mappings for different systems. One can study the structures of Hamiltonians and quantum devices more deeply, there may be more proper parameters of MSP transformation for the systems.

	In the NISQ era, the parameter vector of the MSP transformation is mainly determined by the structure of quantum devices. We would better let the dimension of the parameter vector be equal to the dimension of quantum devices. Moreover, the entries of the parameter vector should be set according to the sizes of quantum devices. For example, if we plan to implement 20-fermion Hamiltonian simulation on a quantum device of size $5 \times 4$, the parameter vector should be set as $\vec{V} = (5, 4)$. However, the entries should be determined by the structure of fermionic systems when the size of quantum devices is large enough. For lattice models, the entries are equal to the corresponding size of the lattice models. For instance, the parameter vector should be $\vec{V} = (3, 4, 5)$ when simulating the Hamiltonian of a lattice model whose size is $3 \times 4 \times 5$ on a three-dimensional quantum device. As for the molecular Hamiltonians, the entries should be set close to $M^{1/D}$, where $D$ is the dimension of quantum devices. $\vec{V}=(10,10)$ will be better than $\vec{V}=(2,50)$ when simulating the Hamiltonian of a 100-orbital molecule.

	With the development of technology, we may be able to build ideal quantum devices in the future. That is, if the qubits are enough and the connection of qubits is all-to-all, then the quantum device is not the determining factor in choosing proper parameter vectors.  We can view the ideal quantum devices as infinite-dimensional systems and thus the dimension of the parameter vector can be set according to the structure of fermionic systems. As we have proved, the Pauli weight of MSP transformation is $\order{L \cdot M^{1/L} }$, where $L$ is the layer of MSP transformation. The Pauli weight increases with $L$, so $L$ should be as large as possible for general systems. While for some systems with special features, it is not the case. We would better utilize these features when setting the parameter vector. For instance, $\vec{V} = (4, 2, 2)$ (SBK) is better than $\vec{V} = (2,2,2,2)$ (BK) for a physical lattice model of size $4 \times 4$. For another example, it would be better to let the first entry of the parameter vector be equal to $5$ if the studied system can be divided into five identical parts. With similar ideas, one can design mappings flexibly based on different qubit configurations and various features of Hamiltonians.

	\section{Numerical test of different mappings}\label{perfermance}
	The Hamiltonian in the second quantization form is given by Eq.~(\ref{Hamiltonian}). We calculate the electronic structure Hamiltonian of $\text{H}_2$ in the minimal basis, STO-3G, and the coefficients are calculated by ChemiQ package with a distance of 0.75000 \AA ~ between two hydrogen atoms,
	\begin{equation}
		\begin{split}
			H= &-1.24728 a_0^\dagger  a_0-0.67284 a_1^\dagger  a_0^\dagger  a_1  a_0\\
			&-0.18177 a_1^\dagger  a_0^\dagger  a_3  a_2-0.48020 a_2^\dagger  a_0^\dagger  a_2  a_0\\
			&-0.66197 a_2^\dagger  a_1^\dagger  a_2  a_1+0.18177 a_2^\dagger  a_1^\dagger  a_3  a_0\\
			&+0.18177 a_3^\dagger  a_0^\dagger  a_2  a_1-0.66197 a_3^\dagger  a_0^\dagger  a_3  a_0\\
			&-0.18177 a_3^\dagger  a_2^\dagger  a_1  a_0-0.69581 a_3^\dagger  a_2^\dagger  a_3  a_2\\
			&-0.48020 a_3^\dagger  a_1^\dagger  a_3  a_1-1.24728 a_1^\dagger  a_1 \\
			&-0.48127a_2^\dagger  a_2-0.48127 a_3^\dagger  a_3.\\
		\end{split}   
	\end{equation}
	Then we apply the JW, BK, and 2SP transformations on the Hamiltonian, and the results are:
	\begin{equation}
		\begin{split}
			H_{\text{JW}}=&-0.81530I -0.04544X_0 X_1 Y_2 Y_3\\
			&+0.04544 X_0 Y_1 Y_2 X_3+0.04544 Y_0 X_1 X_2 Y_3\\
			&-0.04544 Y_0 Y_1 X_2 X_3+0.16988 Z_0+0.16988 Z_1\\
			&-0.21886 Z_2+0.16821 Z_0 Z_1+0.12005 Z_0 Z_2\\
			&-0.21886 Z_3+0.16549 Z_0 Z_3+0.16549 Z_1 Z_2\\
			& +0.12005 Z_1 Z_3+0.17395 Z_2 Z_3,	\\	   
			H_{\text{BK}}=&-0.81530 I+0.04544 X_0 Z_1 X_2\\
			&+0.04544 X_0 Z_1 X_2 Z_3+0.04544 Y_0 Z_1 Y_2 Z_3 \\
			&+0.04544 Y_0 Z_1 Y_2+0.16988 Z_0+0.16988 Z_0 Z_1\\
			&+0.16821 Z_1+0.12005 Z_0 Z_2+0.16549 Z_0 Z_1 Z_2\\
			&-0.21886 Z_2+0.17395 Z_1 Z_3+0.12005 Z_0 Z_2 Z_3\\
			&-0.21886 Z_1 Z_2 Z_3+0.16549 Z_0 Z_1 Z_2 Z_3, \\
			H_{\text{2SP}}=&-0.81530 I+0.04544 X_0 X_2 Z_3\\
			&+0.04544 X_0 Z_1 X_2+0.04544 Y_0 Y_2 Z_3\\
			&+0.04544 Y_0 Z_1 Y_2+0.16988 Z_0+0.16988 Z_0 Z_1\\
			&+0.16821 Z_1+0.12005 Z_0 Z_2-0.21886 Z_2 Z_3\\
			&-0.21886 Z_2+0.17395 Z_3+0.16549 Z_0 Z_1 Z_2\\
			&+0.16549 Z_0 Z_2 Z_3+0.12005 Z_0 Z_1 Z_2 Z_3 .	
		\end{split}
	\end{equation}
	
	We also test other molecules to compare the performance of different mappings. The Pauli weight of parity transformation is much similar to that of JW transformation, and the Pauli weight of SBK transformation and E-type AQM is much larger than that of BK and MSP transformations. So we mainly compare the JW, BK-tree, and MSP transformations. We have compared the performance of the BK and BK-tree transformations on many molecules, and the  BK-tree transformation is found to perform better on the number of mapped Pauli operators and qubits. Hence, it is better to apply the BK-tree transformation instead of the BK transformation. Here, we use the three mappings to transform the Hamiltonians of some molecules and count the number of Pauli operators. The results are shown in Table \ref{tableofPauliWeight}. The MSP transformation always gets a smaller number of Pauli operators than that of the JW and BK-tree transformations.
	\begin{table}[!htbp]
		\centering
		\caption{The numbers of Pauli operators in the electronic structure Hamiltonians.  The parameter vectors of the MSP transformation are in parentheses. All the Hamiltonians are calculated in the STO-3G basis.\label{tableofPauliWeight}}
		\begin{tabular}[c]{llll}
			\hline \hline
			Molecule			&JW			&BK-tree 	 &MSP	\\
			\hline
			H$_2$				&32			& 36		 & 32 (2, 2)		\\	
			LiH					&3888		& 3370		 & 3312 (1,2, 3, 2)	\\
			H$_2$O				&14608		& 12934		 & 12712 (1,2, 3, 3) \\
			NH$_3$				&44708		& 38746		 & 38692 (1,4, 2, 2) \\
			Mg					&17976		& 15414		 & 15270 (1,3, 3, 2)	\\
			N$_2$				&28392		& 23628		 & 22980 (1,5, 2, 2)	\\
			C$_2$H$_2$			&253520		& 197592	 & 191859 (1,3, 2, 3, 2)	\\
			CO$_2$				&428048		& 302926	 & 297930 (1,5, 3, 2)	\\
			C$_2$H$_6$			&1375104	& 934846	 & 932584 (1,4, 2, 2, 2)	\\
			Cl$_2$				&563750		& 369534	 & 366088 (1,3, 3, 2, 2)		\\
			HNO$_3$				&3050496	& 1835853    & 1809180 (1,3, 2, 2, 2, 2)	\\
			CH$_3$COOH			&9305772	& 5374602	 & 5262063 (1,3, 2, 2, 2, 2)	\\	
			CH$_3$COCH$_3$~		&14148158~ 	& 7815190~	 & 7732750 (1,3, 2, 2, 2, 2, 2)	\\
			\hline \hline
		\end{tabular}
	\end{table}
	
	Now we consider the number of gates in the quantum circuit of Hamiltonian simulation. In general, for an item containing $n_x$ Pauli Xs, $n_y$ Pauli Ys, and $n_z$ Pauli Zs, the circuit will require $2(n_x+n_y+n_z-1)$ CNOT gates and $1+2(n_x+n_y)$ single-qubit gates. We count the total number of CNOT and single-qubit gates needed in one Trotter step. The results are shown in Table \ref{tableofGate}. It should be noted that we have ignored the quantum device connection limitations, for analyzing the structure of all the available quantum devices and seeking out the optimum numbers of gates for all of them is hard.
	According to the results, we find that the MSP and BK-tree transformations perform much better than the JW transformation. The MSP transformation with a proper parameter vector performs better than the BK-tree transformation on all the tested results. 
	
	\begin{table}[!htbp]
		\centering
		\caption{The number of gates in the circuit of electronic structure Hamiltonian simulation in one Trotter step. The parameter vectors of the MSP transformation are in parentheses. All the Hamiltonians are calculated in the STO-3G basis.\label{tableofGate}}
		\begin{tabular}[c]{llll}
			\hline \hline
			Molecule			&JW			&BK-tree  	 &MSP	\\
			\hline
			H$_2$				&82			& 74		 & 66 (2, 2)		\\	
			LiH					&10506		& 9822		 & 9258 (1,2, 3, 2)	\\
			H$_2$O				&39435		& 40195		 & 38371 (1,2, 3, 3) \\
			NH$_3$				&120040		& 120524	 & 120368 (1,4, 2, 2) \\
			Mg					&46953		& 47897		 & 46317 (1,3, 3, 2)	\\
			N$_2$				&73802		& 76038		 & 71562 (1,5, 2, 2)	\\
			C$_2$H$_2$			&639080		& 640656	 & 602542 (1,3, 2, 3, 2)	\\
			CO$_2$				&1047183	& 966047	 & 914027 (1,5, 3, 2)	\\
			C$_2$H$_6$			&3338776	& 2956724	 & 2945968 (1,4, 2, 2, 2)	\\
			Cl$_2$				&1348244	& 1194912	 & 1143848 (1,3, 3, 2, 2)	\\
			HNO$_3$				&7130869	& 5992595    & 5834849 (1,3, 2, 2, 2, 2)	\\
			CH$_3$COOH			&21495948	& 17430096	 & 16611850 (1,3, 2, 2, 2, 2)	\\	
			CH$_3$COCH$_3$		&32399164 	& 25346436	 & 24844428 (1,3, 2, 2, 2, 2, 2)\\
			\hline \hline
		\end{tabular}
	\end{table}

	\section{Conclusion and discussion}\label{conclusion}
	In this work, we have presented a theoretical framework, which is useful to discuss many mappings. Furthermore, this framework is a good guide for researchers to design or choose proper mappings for their Hamiltonians and quantum devices. Inspired by the framework, we propose the MSP transformation, which can be reduced to other traditional mappings. In addition, when facing different systems, we can adjust the parameter vector of the MSP transformation to make it more suitable, which is more flexible than other transformations. Finally, we numerically test these mappings and compare their performance on the electronic structure Hamiltonians of various molecules. The MSP transformation performs better than the JW and BK transformations in the Pauli weight and the number of gates.
	
	Indeed, the generation of $S(j)$ is so flexible that $S(j)$ can be generated almost randomly. We should reduce the Pauli weight and improve the contiguity of Pauli strings when designing new mappings, but there is a trade-off between the Pauli weight and the contiguity of Pauli strings. Moreover, mappings also affect the gate cancellation in actual circuits~\cite{tranter2018comparison} and the grouping of Pauli strings in measurements~\cite{hamamura2020efficient}. No single mapping works well for all purposes, so it is important to balance them well. How to achieve the best balance between them is still a challenge, for it depends on the features of Hamiltonians and quantum devices deeply. And the performance of these mappings on real quantum devices also requires further research.
	
	\acknowledgments
	We thank Ye Li for helpful discussion and technical support. The numerical calculations in this paper were performed on the supercomputing system in the Supercomputing Center of University of Science and Technology of China.
	This work was supported by the National Natural Science Foundation of China (Grant No. 12034018) and Innovation Program for Quantum Science and Technology No. 2021ZD0302300.

	\appendix
	\section{The Pauli weight of MSP transformation}\label{Pauli weight of MSP}
	According to Eq. (\ref{operatortrans}), the Pauli weight depends on the size of $P(j)$ and $U(j)$, and actually no more than $|U(j)|+|P(j)|+1$, where $ |B|$ means the size of the set $B$. Note that $P(j)$ and $U(j)$ have no the same elements, thus the worst-case Pauli weight is $|U(j)|+|P(j)|+1=|U(j) \cup P(j)|+1$. We prove that the upper bound of $|U(j) \cup P(j)|$ is $\sum_{l=1}^{L}(v_l-1)$. Let L=1 and $v_1=M$, then the upper bound of $|U(j) \cup P(j)|$ is $M-1=v_1-1$, for there are $M-1$ sites at most except for site $j$. Suppose the upper bound of $U(j)+P(j)$ is $\sum_{l=1}^{L} (v_l-1)$ when the total layer of MSP transformation is $L$. Consider $v_{0}$ the same segments, on which MSP transformation is applied, constitute a larger segment. Compared with the MSP transformation of the $v_0$ segments, the changing of MSP transformation on the larger segment is let the last site of the last segments to store the parity of the total segment. When site $j$ is in the $k$th segment, the changing of $P(j)$ is adding $k-1$ parity sites of the segments which are before the $k$th segment, and the changing of $U(j)$ is adding the last site of the total segment. It should be noted that the last site of the total segment is already in $U(j)$ if site $j$ is in the last segment. Thus there are $v_0-1$ new sites adding to $P(j)+U(j)$ at most. Here we have proved that the upper bound of $|P(j) \cup U(j)|$ is $\sum_{l=1}^{L} (v_l-1$) for any $L$. Therefore, the Pauli weight of the MSP transformation is $(\sum_{l=1}^{L} (v_l-1))+1$.
	
	\section{The algorithm of generating the update, parity, and flip sets}\label{appendix AlgOfThreeSets}
	Given $S(j)$ sets, one can use Alg. \ref{algorithm GeneratingSets} to generate $U(j)$, $P(j)$, and $F(j)$ sets quickly, then it's trival to perform the mapping.  It should be noted that the elements of $S(j)$ should be sorted according to ascending order.
	
	\begin{algorithm}[h]\label{algorithm GeneratingSets}
		\caption{The generating $U(j)$, $P(j)$, and $F(j)$ sets algorithm}
		\KwIn{$M$ and $S(j)$} 
		\KwOut{$U(j)$, $P(j)$, and $F(j)$ }
		\textbf{Initialize}: $U(j)=P(j)=F(j)=\{\}$ \;
		
		\For{$j=0;j<M;j++$}
		{
			\For{$k=j+1;k<M;k++$}{
				
				\If{$j \in S(k)$}{
					\textbf{Pushback} $k$ to $U(j)$    // Generate $U(j)$ \;}
				
			}
		}
		
		\For{$j=0;j<M;j++$}
		{
			\For{$k=0;k<j;k++$}{
				\If{$U(k)=\{\}$ \textbf{or} $U(k)[0] \geq j$}{
					\textbf{Pushback} $k$ to $S(j)$ // Generate $P(j)$  \;}
			}
		}
		
		\For{$j=0;j<M;j++$}
		{
			
			\For{$k$ in $P(j)$}{
				\If{$U(k)[0]= j $}{
					\textbf{Pushback} $k$ to $F(j)$ //Generate $F(j)$ \;}
			}
			
		}

	\end{algorithm}

	\section{The algorithm of realizing MSP transformation}\label{appendix alg}
	We give the algorithm to generate $S(j)$ sets of MSP transformation in Alg. \ref{algorithm MSPT}. 
	\begin{algorithm}[h]\label{algorithm MSPT}
		\caption{The generating $S(j)$ sets of MSP transformation algorithm}
		\KwIn{$M$ and $\vec{V}=(v_1,v_2,\dots,v_L)$}
		\KwOut{$S(j)$}
		
		\textbf{Initialize}: $S(j)=\{\}$ \;
		\textbf{Define Function: Segment($L,R,k$)}\{
		
		\If{$L \neq R$}
		{
			$r=(R-L+1)\mod v_k$ \;
			$u=(R-L+1-r)/v_k$ \;
			\If{$S(R)=\{\}$}
			{
				\For{$j=L;j<R;j++$}{
					
					\textbf{Pushback} $j$ to $S(R)$\;
				}
			}
			\For{$i=1;i\leq v_k, i++$}{
				\If{$r>0$}
				{
					\textbf{Segment}$(L,L+u,k+1)$\;
					$L=L+u+1$\;
					$r=r-1$\;
				}
				\Else{
					\textbf{Segment}$(L,L+u-1,k+1)$ \;
					$L=L+u$\;	
				}

			}
		}
		\Else
		{
			\textbf{Terminate}\;
		}
		\}\;
		
		\textbf{Segment}$(0,M-1,1)$   // Generate $S(j)$ \;

	\end{algorithm}

	\bibliography{MSP}
\end{document}